\documentclass[twoside,twocolumn,9pt]{article}
\usepackage{extsizes,subeqnarray}
\usepackage[super,sort&compress,comma]{natbib} 
\usepackage[version=3]{mhchem}
\usepackage[left=1.5cm, right=1.5cm, top=1.785cm, bottom=2.0cm]{geometry}
\usepackage{balance}
\usepackage{mathptmx,amsmath}
\usepackage{sectsty}
\usepackage{graphicx} 
\usepackage{subfig}
\usepackage{lastpage}
\usepackage[format=plain,justification=justified,singlelinecheck=false,font={stretch=1.125,small,sf},labelfont=bf,labelsep=space]{caption}
\usepackage{float}
\usepackage{fancyhdr}
\usepackage{fnpos}
\usepackage[english]{babel}
\addto{\captionsenglish}{%
  \renewcommand{\refname}{Notes and references}
}
\usepackage{array}
\usepackage{droidsans}
\usepackage{charter}
\usepackage[T1]{fontenc}
\usepackage[usenames,dvipsnames]{xcolor}
\usepackage{setspace}
\usepackage[compact]{titlesec}
\usepackage{hyperref}

\def\l{\left}
\def\r{\right}
\newcommand{\pd}[2]{\frac{\partial #1}{\partial #2}}

\usepackage{epstopdf}
\usepackage[modulo,switch,mathlines]{lineno} 

\definecolor{cream}{RGB}{222,217,201}

\begin{document}

\pagestyle{fancy}
\thispagestyle{plain}
\fancypagestyle{plain}{
\renewcommand{\headrulewidth}{0pt}
}

\makeFNbottom
\makeatletter
\renewcommand\LARGE{\@setfontsize\LARGE{15pt}{17}}
\renewcommand\Large{\@setfontsize\Large{12pt}{14}}
\renewcommand\large{\@setfontsize\large{10pt}{12}}
\renewcommand\footnotesize{\@setfontsize\footnotesize{7pt}{10}}
\makeatother
\def\v{\vspace{2cm} }
\renewcommand{\thefootnote}{\fnsymbol{footnote}}
\renewcommand\footnoterule{\vspace*{1pt}%
\color{cream}\hrule width 3.5in height 0.4pt \color{black}\vspace*{5pt}} 
\setcounter{secnumdepth}{5}

\makeatletter 
\renewcommand\@biblabel[1]{#1}            
\renewcommand\@makefntext[1]%
{\noindent\makebox[0pt][r]{\@thefnmark\,}#1}
\makeatother 
\renewcommand{\figurename}{\small{Fig.}~}
\sectionfont{\sffamily\Large}
\subsectionfont{\normalsize}
\subsubsectionfont{\bf}
\setstretch{1.125} 
\setlength{\skip\footins}{0.8cm}
\setlength{\footnotesep}{0.25cm}
\setlength{\jot}{10pt}
\titlespacing*{\section}{0pt}{4pt}{4pt}
\titlespacing*{\subsection}{0pt}{15pt}{1pt}

\fancyfoot{}
\fancyhead{}
\renewcommand{\headrulewidth}{0pt} 
\renewcommand{\footrulewidth}{0pt}
\setlength{\arrayrulewidth}{1pt}
\setlength{\columnsep}{6.5mm}
\setlength\bibsep{1pt}

\makeatletter 
\newlength{\figrulesep} 
\setlength{\figrulesep}{0.5\textfloatsep} 

\newcommand{\topfigrule}{\vspace*{-1pt}%
\noindent{\color{cream}\rule[-\figrulesep]{\columnwidth}{1.5pt}} }

\newcommand{\botfigrule}{\vspace*{-2pt}%
\noindent{\color{cream}\rule[\figrulesep]{\columnwidth}{1.5pt}} }

\newcommand{\dblfigrule}{\vspace*{-1pt}%
\noindent{\color{cream}\rule[-\figrulesep]{\textwidth}{1.5pt}} }
 \DeclareGraphicsExtensions{.pdf,.jpeg,.png,.jpg,.eps}

\makeatletter
\def\linenumberfont{\normalfont\tiny\sffamily} 
\let\linenumberparskip\@empty
\def\linebox#1{\hbox to\textwidth{%
  \if@twocolumn
    \ifnum\column@position=1 
      \rlap{\hskip-\linenumbersep\linenumberfont #1}%
    \else 
      \llap{\hskip\textwidth\linenumberfont #1}%
    \fi
  \else 
    \rlap{\hskip-\linenumbersep\linenumberfont #1}%
  \fi}}
\makeatother

\makeatother

\twocolumn[
  \begin{@twocolumnfalse}
\vspace{1em}
\sffamily
\begin{tabular}{m{4.5cm} p{13.5cm} }

 & \noindent\LARGE{\textbf{Designing optimal  elastic filaments for viscous propulsion}} \\
\vspace{0.3cm} & \vspace{0.3cm} \\

 & \noindent\large{Mariia Dvoriashyna$^{a,b}$ and Eric Lauga$^a$} \\

  & \noindent\normalsize{The propulsion of many eukaryotic cells is generated by flagella, flexible slender filaments that are actively oscillating in space and time. The dynamics of these biological appendages have inspired the design of many types of artificial microswimmers. The magnitude of the filament's viscous propulsion depends on the time-varying shape of the filament, and that shape depends in turn on the spatial distribution of the bending rigidity of the filament. In this work, we rigorously determine the relationship between the mechanical (bending) properties of the filament and the viscous thrust it produces using mathematical optimisation. Specifically, by considering a model system (a slender elastic filament with an oscillating slope at its base), we derive the optimal bending rigidity function along the filament that maximises the time-averaged thrust produced by the actuated filament. Instead of prescribing a specific functional form, we use functional optimisation and adjoint-based variational calculus to formally establish the link between the distribution of bending rigidity and propulsion. The optimal rigidities are found to be stiff near the base, and soft near the distal end, with a spatial distribution that depends critically on the constraints used in the optimisation procedure. These findings may guide the optimal design of future artificial swimmers.}


\end{tabular}

 \end{@twocolumnfalse} \vspace{0.6cm}

  ]

\renewcommand*\rmdefault{bch}\normalfont\upshape
\rmfamily
\section*{}
\vspace{-1cm}


\footnotetext{$^a$Department of Applied Mathematics and Theoretical Physics, University of Cambridge, Wilberforce Rd, Cambridge CB3 0WA, UK \\
$^b$School of Mathematics and Maxwell Institute for Mathematical Sciences, University of Edinburgh, Peter Guthrie Tait Rd, Edinburgh EH9 3FD, UK
\\ 
Email: m.dvoriashyna@ed.ac.uk, e.lauga@damtp.cam.ac.uk}





\section{Introduction}

The physics of motile microorganisms, such as bacteria, microalgae and spermatozoa has recently been a subject of active research at the intersection between physics, mathematics and biology.
\cite{fauci2006biofluidmechanics,koch2011collective,goldstein2015green,wadhwa2022bacterial,lauga2009hydrodynamics}
Studying cell motility  is  important not only for our general understanding of biological and biophysical phenomena,\cite{gaffney2011mammalian,lauga2016bacterial} but also for the development of biomimetic, synthetic microrobots in a myriad of potential applications, including drug delivery, smart surgery, sensing and detoxification.\cite{nelson2010microrobots,peyer2012bacteria,peyer2013bio,li2017micro} 

Inspired by the actuation and geometry of motile microorganisms, different designs of swimming microrobots have   been developed,\cite{chen2017recent,zhou2021magnetically} coming in a variety of  shapes (e.g.~rigid helical filaments, straight or helical elastic filaments, collection of beads), actuation mechanisms (e.g.~rotation of a helical shape, planar or circular oscillation of an elastic tail, prescribed shape changes) and materials (cobalt composite on helical lipids, nickel or iron-filled carbon nanotubes, magnetic beads linked with DNA strands).\cite{peyer2013bio,schuerle2012helical,dreyfus2005microscopic,sun2020magnetically,jang2015undulatory,zhang2009artificial} For example, recently developed undulatory multilink nano-swimmers are controlled by a planar oscillating magnetic field, which sets  elastic tails into oscillatory motion and creates forward propulsion, similar to the swimming of spermatozoa.\cite{jang2015undulatory} Another example includes artificial bacterial flagella, in which  a rigid helical shape  attached to a thin soft-magnetic head rotated by an external rotating magnetic field, resulting in a corkscrew-type propulsion similar to that of flagellated bacteria. \cite{zhang2009artificial}

Biological microorganisms swim on small scales, with relevant Reynolds numbers ranging between $\mathcal{O}(10^{-5}) $ and $\mathcal{O}(10^{-1})$.\cite{lauga2020fluid} In this regime, inertia of both the swimmer and the surrounding fluid are negligible, and swimming strategies have to rely on viscous (Stokesian) forces to generate propulsion. One of the simplest swimming modes capable of creating propulsive forces at zero Reynolds number is the 
 classical `flexible oar', initially proposed  by  Purcell in his famous lecture on `life at low Reynolds numbers'. \cite{purcell1977life} In this setup, a flexible tail is made to passively oscillate in a viscous fluid, generating travelling waves along the filament (whose amplitude tends to decay exponentially if the forcing is localised along the filament) and resulting in propulsion in the  direction opposite to that of the wave propagation.\cite{machin1958wave,wiggins1998flexive}

 This mode of swimming has  been used experimentally in multiple  artificial swimmers. For instance, \citeauthor{guo2008development}\cite{guo2008development} developed a (cm)-scale fish-like microbot, with a main body made of wooden and styrol materials placed on a permanent magnet, and the fin made of a polyimide  film sheet. Directly inspired by spermatozoa, \citeauthor{williams2014self}\cite{williams2014self} developed a (mm)-scale biohydrid swimmer  made of PDMS and actuated by contractile cardiomyocyte cells lining the base of the filament. \citeauthor{magdanz2020ironsperm}
  \cite{magdanz2020ironsperm,magdanz2022ironsperm} developed IRONSperm, which comprised of bovine sperm cells ($\approx$60 $\mu$m long) coated with a suspension of 100-nm rice grain-shaped maghemite nanoparticles and controlled by a magnetic field. A related device was designed by \citeauthor{celi2021artificial}\cite{celi2021artificial}, composed of a superparamagnetic head and a flexible gold/polypyrrole (Au/PPy) flagellum, and actuated by a magnetic field. \citeauthor{mushtaq2019motile}\cite{mushtaq2019motile} developed a nano-eel, a multifunctional piezoelectric tailed nanorobot, which consisted of a smart flexible tail made of a polyvinylidene fluoride-based copolymer linked to a PPy nanowire head, decorated with nickel (Ni) rings for magnetic actuation. 
   Theoretical modelling can help understanding physics underlying propulsion of these artificial microswimmers and propose ways of enhancing it.

From a theoretical standpoint, and following the pioneering contributions of Machin,\cite{machin1958wave} the dynamics and propulsion of an elastic filament with constant thickness oscillating in a viscous fluid were   modelled physically and mathematically in classical studies. \cite{wiggins1998flexive,wiggins1998trapping}  These original  models relied on a small-amplitude approximation, which allows linearisation of the equations of motion for the shape of the elastic filament. The propulsion is generated as a result  of the quasi-steady balance of viscous and elastic forces. These linear models have been validated experimentally, using a (cm)-scale robot with an elastic tail actuated at its base, immersed in high viscosity silicone oil.\cite{yu2006experimental}  
Beyond the small-amplitude approach, the full elastohydrodynamic problem of an oscillating filament was studied numerically using the lattice-Boltzmann method, \cite{wu2014simulation} asymptotic coarse-grained elastohydrodynamics, \cite{moreau2018asymptotic} and regularised Stokeslets, \cite{jabbarzadeh2020numerical} while further work also addressed the non-linear dynamics of biological filaments resulting  from their molecular structure. \cite{hilfinger2009nonlinear,oriola2017nonlinear}

With a view on applications in micro-scale robotics, designing a hydrodynamically optimal synthetic swimmer, i.e.~one that can self-propel with the highest swimming velocity for a given energy expenditure,  is a problem of fundamental interest.  Biologically, this optimisation problem has received a lot of attention in the  context of motile eukaryotic cells. It was shown that  spermatozoa have an optimal flagellum-to-head length ratio, with the most efficient swimming mode involving symmetrical travelling waves in the flagellum.\cite{tam2011optimal}  The travelling waves were shown to lead to the optimal motion in an active filament, \cite{pironneau1974optimal,pironneau1975optimal,lauga2020traveling} while the mathematically optimal shape of the wave was shown to be sawtooth-like\cite{lighthill1975mathematical}, with shape singularities that may be regularised by  additional biophysical constraints.\cite{spagnolie2010optimal,lauga2013shape} 
Recently,  by making  use of the  new BOSO-Micro database,\cite{velho2021bank} which includes a comprehensive collection of  experimental data from the literature  on the swimming speed and morphological characteristics of unicellular organisms, it was shown  that amplitude-to-wavelength aspect ratios of flagellar eukaryotes are very close to the hydrodynamic optimum.\cite{lisicki2024eukaryotic}  Additionally, optimal head-to-tail ratios were studied for a model swimmer equipped with elastic filament actuated at the base,\cite{lauga2007floppy} and the optimal shape of the head was investigated for a synthetic swimmer with magnetic head and an elastic tail.\cite{gadelha2013optimal}

\begin{figure}
\includegraphics[width=0.5\textwidth]{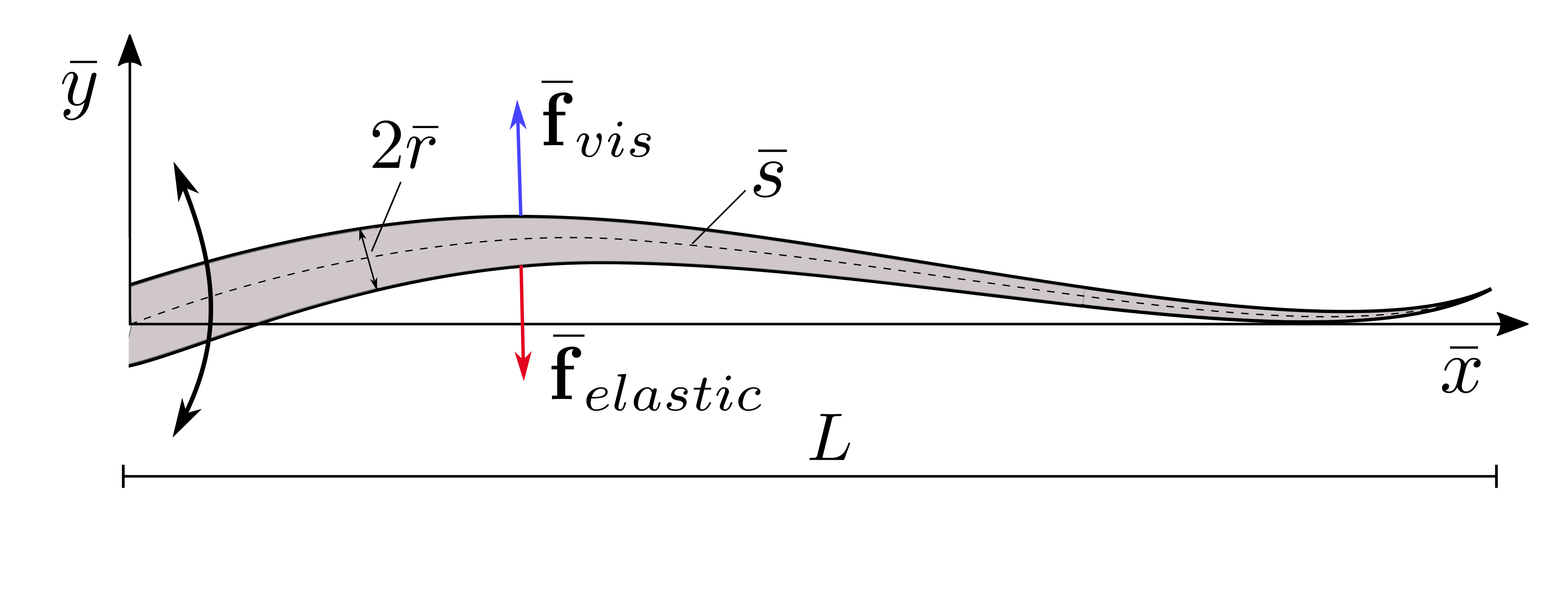}
\caption{Sketch of the model problem optimised in this paper: An elastic filament of variable thickness is made to oscillate about its hinged base. At each instant, its shape is obtained as the balance between the viscous forces ($\mathbf{\bar{f}}_{vis}$) and the elastic forces ($\mathbf{\bar{f}}_{elastic}$). We use variational calculus, along with an adjoint field, to determine the bending rigidity  leading to maximal propulsion.}\label{fig:sketch1}
\end{figure}

Focusing on the specific case of actuated elastic filaments, past work demonstrated that tapering an elastic filament may lead to an increase in viscous propulsion.\cite{rathore2012planar} For a filament with a circular cross-section of radius $r(s)$, which depends on the arc-length $s$ along the filament (see e.g.~Fig.~\ref{fig:sketch1}), the bending rigidity of the filament, $A$,  scales as $A\sim r^4$. Thus, the tapered filaments are  more rigid near the actuation point than the free end. Further studies considered specific functional forms for the filament radius $r$ or the bending rigidity $A$, such as linear, quadratic, and exponential functions of $s$. \cite{rathore2012planar,singh2017effect,kotesa2013tapered,peng2017maximizing} In most cases, the propulsive force was shown to increase if the filament is stiffer at the base and softer at the distal end. A recent experimental investigation suggested that a microrobot equipped with DNA-based filament wrapped in a tapered bundle swims faster than the one with a bundle of constant thickness.\cite{maier2016magnetic} These studies reveal  a nontrivial relationship between the bending rigidity of the filament (in particular, how it cross-section varies) and the magnitude of the propulsion generated.

In this work, we propose to rigorously determine the relationship between the mechanical properties of the filament and  the produced thrust. By considering a specific model system consisting of a beating elastic filament with an oscillating slope at its base, we determine the optimal bending rigidity function  that maximises the time-averaged thrust produced by the actuated filament. Instead of prescribing a specific functional shape, we use functional optimisation and variational calculus to establish formally the link  between the bending rigidity function and propulsion for different sets of constraints. We find that the optimal bending rigidity for this model swimmer is invariably stiff at the base of the filament and soft at the distal end.

This paper is structured as follows. We first formulate the model problem of a beating elastic filament with an oscillating slope at its base in \S\ref{sec:modelsystem}, and simplify it in \S\ref{sec:simplification}. We then pose in \S\ref{sec:Optimisation}  the optimisation problem consisting in finding the cross-sectional shape that  maximises the propulsive force generated by the filament. It is solved using an adjoint approach, which is  implemented numerically in a steepest ascent algorithm  in \S~\ref{sec:algo}.  We present the results of the optimisation procedure in \S~\ref{sec:res1} and \S~\ref{sec:res2} and conclude with a discussion  in \S~\ref{sec:disc}.

\section{Model, Methods and Optimisation}
\label{sec:methods}
\subsection{Model system: Oscillating slender elastic filament}\label{sec:modelsystem}

We study in this paper a well-defined model system, which will allow us to solve the optimisation problem rigorously.

\subsubsection{Geometry}
We consider the case of an oscillating slender elastic filament   whose slope at its base is made to oscillate in time at a prescribed frequency (see sketch in  Fig.~\ref{fig:sketch1}). Using bars to denote dimensional variables, the filament is aligned on average with the   $\bar{x}$ axis and its base is  located at  $\bar{x}=0$.  With $\bar{s}$ denoting the arclength along the filament, oscillations of the slope of the filament at  $\bar{x}=\bar{s}=0$ result in a planar wave propagating between the filament base and its free end located at $\bar{s}=L$.  We denote the  position of the centreline of the filament at time $\bar{t}$ by $\bar{y}(\bar{s},\bar{t})$, while the circular cross-section  of the  filament has radius $\bar{r}(\bar{s})$. The filament is assumed to be slender, so $\bar{r}(\bar{s})\ll L$ everywhere. The goal of the paper is to determine  the function $\bar{r}(\bar{s})$ which maximises the propulsion generated by the filament.

\subsubsection{Equations of motion}\label{sec:eqs}

In order to tackle the problem analytically, we consider a linearised version of the elastohydrodynamic problem, assuming that the displacements of the filament are always small, so that $\bar{s}\approx \bar{x}$. This approach captures the physics of the problem, and is known to remaining quantitatively accurate even beyond the strict small-amplitude regime of validity.\cite{yu2006experimental}

In the absence of inertia, the shape of the filament is determined by a quasi-steady balance between hydrodynamic and elastic forces. To compute the viscous forces, we exploit the slenderness of the filament  and  use resistive force theory (RFT), which  relates the hydrodynamic force density at any point along the filament to its local velocity via  drag coefficients in the directions orthogonal and parallel to the local tangent to the filament,  $\zeta_{\perp}=4\pi \mu / (\ln (L/\bar{r})+1/2)$, $\zeta_{\parallel} = 2\pi \mu / (\ln (L/\bar{r})-1/2)$,\cite{cox1970motion} where $\mu$ denotes the  dynamic viscosity of the fluid. Within the assumptions of RFT, the viscous force density per unit length of the filament at each point   can be written as \begin{equation}\label{eq:RFT}
\mathbf{\bar{f}}_{vis} = -\zeta_{\perp} \mathbf{\bar{v}} - (\zeta_{\parallel} -\zeta_{\perp})\mathbf{t}(\mathbf{t}\cdot \mathbf{\bar{v}}),
\end{equation}
 where $\mathbf{t}$ is the local tangent vector, and $\mathbf{\bar{v}}$ is the local velocity distribution along the swimmer.\cite{lauga2020fluid} For small-amplitude beating, the viscous force occurs primarily in the direction orthogonal to the flagellum axis (i.e.~the $\bar{y}$ direction), with density 
 \begin{equation}\label{f:visc}
\bar{f}_{vis}=-\zeta_{\perp}\bar{v},
\end{equation}
 where $\bar{v}=\pd{\bar{y}}{\bar{t}}$ is the instantaneous velocity in $\bar{y}$  direction.

To compute the elastic force density, $\mathbf{\bar{f}}_{elastic}$, we make use of  classical elastic beam theory.\cite{wiggins1998flexive,ochsner2021classical} For displacement of the filament  written as $\bar{y}(\bar{x})$, the leading-order elastic force in the $\bar{y}$ direction is   given by
\begin{equation}\label{f:ela}
\bar{f}_{elastic}=-\pd{^2}{\bar{x}^2}\l(\bar{A}(\bar{x})\pd{^2 \bar{y}}{\bar{x}^{2}}\r),
\end{equation}
 where $\bar{A}(\bar{x})=E\bar{I}(\bar{x})$ is the bending stiffness, a product of  the material's Young's modulus, $E$, (assumed to be constant) and the second moment of cross-sectional area, $\bar{I}(\bar{x})=\pi \bar{r}(\bar{x})^4$. 
 
 Imposing a quasi-steady balance between the viscous and elastic forces,  equations~\eqref{f:visc}-\eqref{f:ela}, we obtain the governing (classical) hyper-diffusion equation that describes the position of the centreline of the filament in time, in the linearised limit, as
 \begin{equation}
 \zeta_\perp \pd{\bar{y}}{\bar{t}}=-\pd{^2}{\bar{x}^2}\l(\bar{A}(\bar{x})\pd{^2\bar{y}}{\bar{x}^2}\r).\label{eq:force_balance}
 \end{equation}

Note that the inextensibility of the filament comes in at higher order in the oscillation amplitude and, hence, does not appear in the linearised version above. 

\subsubsection{Boundary conditions}\label{sec:bcs}

In the specific model problem considered in this paper, we  prescribe that the  slope of the filament at its base  oscillates  with  frequency $\omega$ and impose that the position of the filament is fixed at the base. At the free end, we adopt force and torque-free conditions. The four boundary conditions (BCs) therefore read
\begin{subequations}\label{eq:BCs}
\begin{align}
& \bar{y}(0,\bar{t})=0  &&\qquad \mbox{fixed base,} \label{eq:BC1}\\
& \pd{\bar{y}}{\bar{x}}(0,\bar{t}) =\varepsilon \sin \omega \bar{t} &&\qquad \mbox{oscillation of the slope,} \label{eq:BC2}\\
& \pd{^2 \bar{y}}{\bar{x}^2} (L,\bar{t})=0 &&\qquad \mbox{torque free end,}\\
& \pd{^3\bar{y}}{\bar{x}^3}(L,\bar{t}) =0&&\qquad \mbox{force free end}.
\end{align}
\end{subequations}

Therefore, in the linearised limit $\varepsilon \ll 1$, the leading-order position of the  filament centreline satisfies equation \eqref{eq:force_balance} along with the BCs in equation \eqref{eq:BCs}.

 \subsubsection{Propulsive force}
 
In order to compute the propulsive force, we need to calculate the component of the viscous force experienced by the filament in the $\bar x$ direction, i.e.~perpendicular to the beating direction. 
Although the  drag coefficients  $\zeta_\perp$ and  $\zeta_\parallel$ appearing in equation~\eqref{eq:RFT} depend on the local cross section of the filament, their dependence is logarithmic; we may thus treat them as constant in the slender limit.

 Within the small-amplitude approximation we can write $\mathbf{t}\approx (1,\partial \bar{y}/\partial \bar{x})$ and $\mathbf{\bar{v}} = (0,\partial \bar{y}/\partial \bar{t})$. The local force in the $\bar{x}$ direction is then classically given by $\bar{f}_{x} = -(\zeta_{\parallel} -\zeta_{\perp}) \frac{\partial \bar{y}}{\partial \bar{t}}\frac{\partial\bar{y}}{\partial \bar{x}}$ (please see Refs.~\cite{pak2015theoretical,lauga2020fluid} for detailed derivation).  Since waves are expected to propagate in the positive $\bar{x}$ direction, and propulsion to occur in the opposite direction,  the  magnitude of the time-averaged propulsive force  experienced by the filament,  $\bar{F}$,  is  obtained   by taking a double integral in time and space of $-\bar{f}_{x}$, so that
 \begin{equation}\label{eq:F_dim}
 \bar{F} = -\frac{\omega}{2\pi} \int_0^{2\pi/\omega}\int_0^L (\zeta_{\perp} -\zeta_{\parallel})\pd{\bar{y}}{\bar{t}}\pd{\bar{y}}{\bar{x}} {\rm d}\bar{x} {\rm d}\bar{t}.
 \end{equation}
 With this sign convention, the magnitude of the force is positive (and the net force acts  in the negative $\bar{x}$ direction).

\subsection{Model simplification}
\label{sec:simplification}
\subsubsection{Non-dimensionalisation}
The first step in  simplifying the model consists in non-dimensionalising the problem. The relevant time scale is given by the inverse frequency of oscillations, while the relevant length scales are the length of the filament $L$ along the $\bar{x}$ direction, and the small-amplitude motion of magnitude $\varepsilon L$ along $\bar{y}$. A relevant scale $A_0$ for the rigidity  can be defined using the length of the filament, $A_0=L^4 \omega \zeta_{\perp}$, so that $\bar{A}=A_0A$. This choice  corresponds to a relevant dimensionless Sperm number, defined  as $Sp = L\l(\zeta_{\perp}\omega/A_0A\r)^{1/4}$, of one when $A=1$.\cite{lauga2020fluid}

 We thus define the dimensionless variables
 \begin{equation}\label{eq:scales}
 y=\bar{y}/\varepsilon L, \qquad t=\bar{t}\omega, \qquad x=\bar{x}/L,\qquad  A=\bar{A}/A_0,
 \end{equation}
 and the dimensionless version of \eqref{eq:force_balance} is now
  \begin{equation}
\pd{y}{t}=-\pd{^2}{x^{2}}\l(A(x)\pd{^2y}{x^{2}}\r).\label{eq:force_balance_dimless}
 \end{equation}
 The time averaged propulsive force   $ \bar{F} $ in  equation~\eqref{eq:F_dim} now scales with relevant magnitude  $(\zeta_\perp-\zeta_\parallel)\varepsilon^2 L^2 \omega$ and the  dimensionless  force, $F$, is given by 
  \begin{equation}\label{eq:F_prop_y}
F = -\frac{1}{2\pi}\int_0^{2\pi}\int_0^{1}\pd{y}{t}\pd{y}{x}{\rm d}x{\rm d}t.
 \end{equation}

\subsubsection{Separation of variables}
 To further simplify the mathematical problem, we exploit the linear dynamics and write the time-varying amplitude of the dimensionless filament as
\begin{equation}\label{eq:sep}
y(x,t)=f(x)\cos t+g(x)\sin t.
\end{equation}
Substituting the expression for $y$ in equation~\eqref{eq:sep} 
into  the force balance in equation \eqref{eq:force_balance_dimless}, we obtain two coupled equations for $f(x)$ and $g(x)$ as
\begin{subeqnarray}\label{eq:no_time_problem}
\slabel{eq:12a}  - f+\l(A(x)g''\r)''&=&0,\\ 
\slabel{eq:12b}   g+\l(A(x)f''\r)''&=&0.
\end{subeqnarray}
We will denote with $\mathcal{C}_1$ and $\mathcal{C}_2$ the left hand sides of   equations~\eqref{eq:12a} and \eqref{eq:12b}, respectively. 

Using equation~\eqref{eq:sep},  the propulsive force from equation \eqref{eq:F_prop_y} can be then simplified as
\begin{eqnarray}\label{eq:fprop_fg}
&& F = -\frac{1}{2\pi} \int_0^{2\pi}\frac{1}{2}\int_0^{1}\l[(gg'-ff')\sin 2t+f'g\cos^2 t-g'f \sin^2 t\r]{\rm d}x{\rm d}t \nonumber \\
&& \qquad \qquad =-\frac{1}{2} \int_0^{1}(f'g-g'f){\rm d}x.
\end{eqnarray}

Finally,  the non-dimensional boundary conditions \eqref{eq:BCs}   transform into  
\begin{subeqnarray}\label{eq:no_time_problem-bc}
    &f(0)=f'(0)=f''(1)=f'''(1)=0, \slabel{eq:no_time_problem-bc-1} \\
    &g'(0)=1,\ g(0)=g''(1)=g'''(1)=0.\slabel{eq:no_time_problem-bc-2}
 \end{subeqnarray}

\subsection{Optimisation of bending rigidity}
\label{sec:Optimisation}

In this section we outline the mathematical formalism used to determine the bending rigidity of the filament leading to maximal  propulsion. 

 \subsubsection{Dimensional optimisation problem}\label{sec:opt}
The dimensional optimisation problem consists in maximising the magnitude of the time-averaged propulsive force  $\bar{F}$ (which is positive) over all possible distributions of bending rigidities, $\bar{A}$,  subject to the equations for the filament dynamics, i.e.
\begin{equation}\label{eq:the_vaiation_problem_dim}
\max_{\bar{A}} \bar{F} \quad  \mbox{subject to}\ \eqref{eq:force_balance},\eqref{eq:BCs}.
\end{equation}

\subsubsection{Dimensionless optimisation problem}
In dimensionless variables, the optimisation problem in equation~\eqref{eq:the_vaiation_problem_dim} can be rewritten as
\begin{equation}\label{eq:the_vaiation_problem}
\max_{A} F \quad  \mbox{subject to}\ \mathcal{C}_1=0 \,\eqref{eq:12a}, \ \mathcal{C}_2=0\, \eqref{eq:12b},\ \mbox{and BCs \eqref{eq:no_time_problem-bc}}.
\end{equation}

The functional optimisation problem in equation~\eqref{eq:the_vaiation_problem} in general needs to be solved numerically.

\begin{figure}[t]
\includegraphics[width=0.5\textwidth]{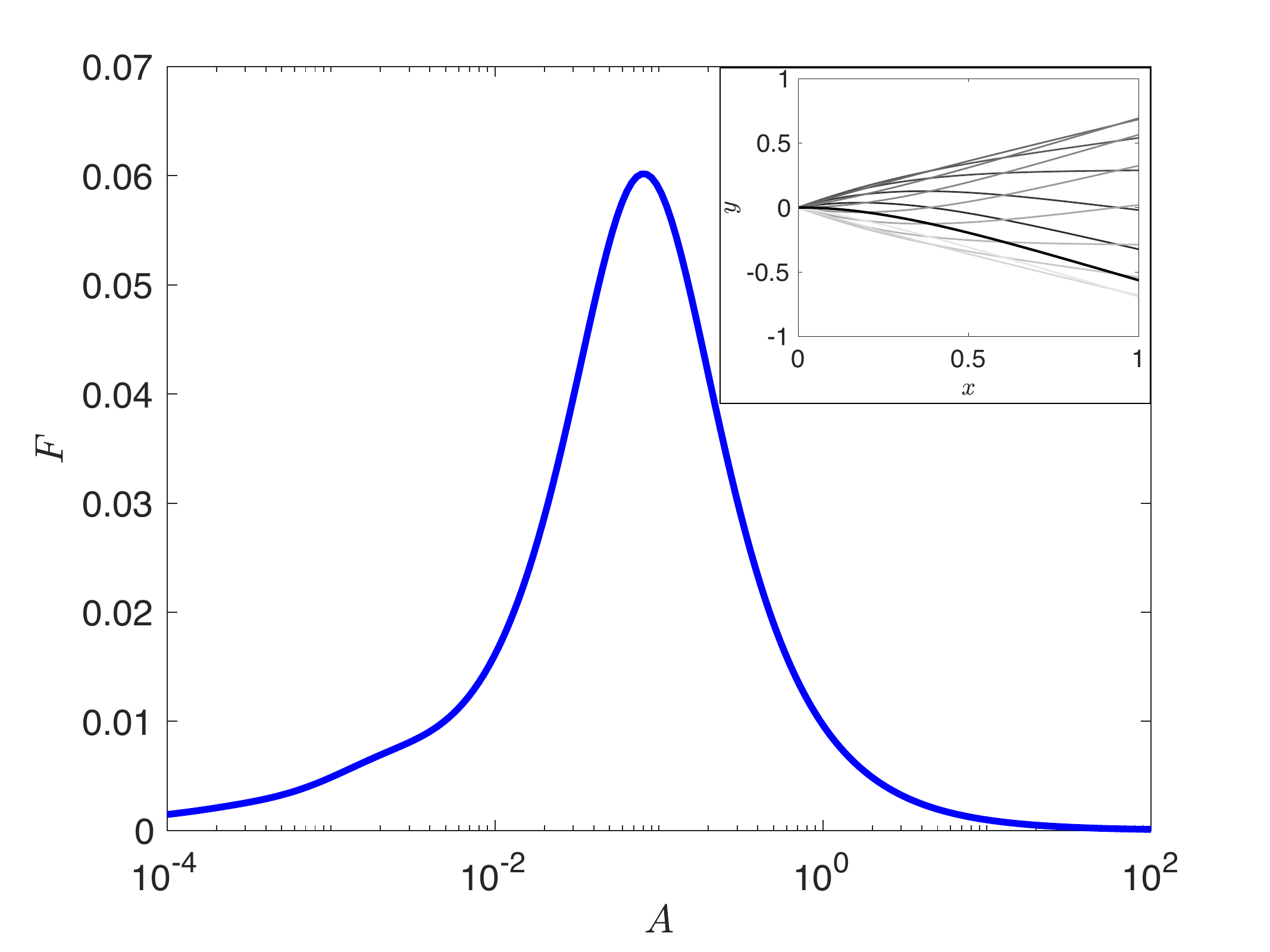}
\caption{Time-averaged (dimensionless) propulsive force in the case of a constant bending rigidity ($A$).  Maximum propulsion for a hinged filament is  $F\approx 0.06$ at the optimal value $A\approx 0.0794$. Inset: time-varying position of the hinged filament $y(x)$ at different points in time for this optimal value of $A$: $t=0$ is depicted with a thick black line and the following time steps are shown with decreasing greyscale. }\label{fig:const-A}
\end{figure}

\subsubsection{Special case: Constant bending rigidity}

If we consider the special case of  a spatially homogeneous $A$, the optimisation  can be carried out analytically, as shown in previous work.\cite{wiggins1998flexive} In Fig.~\ref{fig:const-A}, we plot the dependence of the resulting dimensionless propulsive force on the value of the bending rigidity ($A$). There is almost no propulsion for very flexible (small $A$) or very rigid filaments (large $A$), and an optimal value is $A\approx 0.0794$. In the inset we also display the time-varying position of  the centreline of the filament, $y(x,t)$, for   this optimal choice  of $A$.

\subsubsection{Functional derivative and adjoint functions}\label{sec:fun_der}

To solve the functional optimisation problem~\eqref{eq:the_vaiation_problem} without introducing any assumptions on the specific shape of $A$, we will use variational calculus and will compute the functional derivative of ${\rm d}F/{\rm d}A$ with the help of two adjoint functions.

To derive the adjoint problem we consider small perturbations, $A \to A+ \delta A$, $f \to f +\delta f$ and $g \to g +\delta g$, of both $\mathcal{C}_1$ \eqref{eq:12a} and $\mathcal{C}_2$ \eqref{eq:12b}. Keeping only the first-order terms we obtain
 \begin{subeqnarray}
\slabel{eq:pert_f} \delta f - (\delta A g'')'' - (A\delta g'')''=0,\\
 \delta g + (\delta A f'')'' + (A \delta f'')''=0. \slabel{eq:pert_g}
 \end{subeqnarray}
We next write equation \eqref{eq:pert_f}   in weak form, multiplying by a test function $h$ and integrating over the domain
\begin{eqnarray}
\int_0^{1} (\delta f - (A\delta g'')'')h {\rm d}x =\int_0^{1}  (\delta A g'')''h {\rm d}x.
\end{eqnarray}
Integrating by parts four times   the second term of the left-hand side, and two times  the right-hand side, and applying boundary conditions \eqref{eq:no_time_problem-bc} leads to
\begin{eqnarray}
\int_0^{1} (h\delta f - (Ah'')''\delta g ){\rm d}x + (A\delta g'')'h(0)-A\delta g'' h'(0)-\nonumber \\
\qquad \qquad  -A\delta g'h''(1)+\delta g (h'' A)'(1) =\int_0^{1}  \delta A g''h''{\rm d}x. \label{eq:adj-int-c1}
\end{eqnarray}
 We will specify the boundary terms later. 
 
 Following a similar calculation, the weak form of equation \eqref{eq:pert_g} with a test function $j$ can be written as
 \begin{eqnarray}
\int_0^{1} (j\delta g + (Aj'')''\delta f){\rm d}x -(A\delta f'')'j(0)+A\delta f'' j'(0)+ \nonumber \\
\qquad +A\delta f'j''(1)-\delta f (j''A)'(1)=-\int_0^{1}  \delta A f''j''{\rm d}x .\label{eq:adj-int-c2}
\end{eqnarray}

Taking the difference between equation~\eqref{eq:adj-int-c1} and equation~\eqref{eq:adj-int-c2} we obtain  
\begin{eqnarray}
&& \int_0^{1}  [h-(Aj'')'']\delta f - [j+(A h'')'']\delta g {\rm d}x +(A \delta g'')'h(0)+\nonumber \\
&& (A\delta f'')'j(0)-A\delta g'' h'(0)-A\delta f'' j'(0)-A\delta g'h''(1)-A\delta f'j''(1)+\nonumber \\ 
&& \delta g (h'' A)'(1)+\delta f (j'' A)'(1) =\int_0^{1}  \delta A (f''j''+g''h''){\rm d}x.\label{eq:adj-der}
\end{eqnarray}

The variation of the propulsive force $F\to F+\delta F$ in equation~\eqref{eq:fprop_fg}  at  linear order reads
\begin{equation}
 \delta F = -\frac{1}{2}\int_0^1(f'\delta g+ \delta f' g - g'\delta f-\delta g' f){\rm d}x,
\end{equation}
which after   integration by parts becomes 
\begin{equation}
 \delta F = -\frac{1}{2} \l[g(1)\delta f(1) -\delta g(1) f(1)+2\int_0^1 ( \delta g f' - g'\delta f) {\rm d}x \r].\label{eq:force-der}
\end{equation}

In order to compute the functional derivative, we need to make an explicit link between $\delta A$ and $\delta F$. To do this, we equate the left-hand side of \eqref{eq:adj-der} with the right-hand side of~\eqref{eq:force-der}, by choosing the adjoint functions  $h$ and $j$ to satisfy the  field equations
\begin{subeqnarray}\label{eq:adj-problem}
h-(Aj'')''&=&g',\slabel{eq:adj-c1}\\
j+(Ah'')''&=&f',\slabel{eq:adj-c2}
\end{subeqnarray}
associated with  the boundary conditions
\begin{subeqnarray}\label{eq:adj-bc}
h(0)=h'(0)=h''(1)=0, && (Ah'')'(1)=f(1)/2,\\
j(0)=j'(0)=j''(1)=0, && (Aj'')'(1)=-g(1)/2.
\end{subeqnarray}
With this choice of $h$ and $j$, we obtain an explicit equation linking  $\delta F$ with $\delta A$ and equation~\eqref{eq:force-der}   becomes
\begin{eqnarray}
\delta F =\int_0^{1}  \delta A (f''j''+g''h''){\rm d}x.
\end{eqnarray}
 The functional derivative of the propulsion $F$ in the space of $f$ and $g$ functions is given by
\begin{eqnarray}\label{eq:func-der}
{\rm d} F/{\rm d} A:=f''j''+g''h'',
\end{eqnarray} 
with the four functions satisfying equations  \eqref{eq:no_time_problem} (physical problem) and \eqref{eq:adj-problem} (adjoint problem)  with boundary conditions in  equations  \eqref{eq:no_time_problem-bc} and \eqref{eq:adj-bc}.

\subsection{Constraints on bending and numerical algorithm}\label{sec:algo}

\subsubsection{Singular solutions}

To compute the solution to the optimisation problem, we use the functional derivative obtained in 
equation \eqref{eq:func-der} and the analogue of steepest ascent in   functional space to find the optimal solution. In other words, we set up an iteration procedure, and at every iteration step follow the functional derivative to increase the value of $F$.  \cite{hinze2008optimization} Without additional constraints, however,  computations show that the solution always ends up being singular, with the bending rigidity (and hence the thickness of the filament) becoming zero at some point along the filament. Note that for sufficiently soft filaments (high Sp numbers), the assumptions of our linearised model may become invalid. Thus, additional constraints must be considered to regularise the problem and to keep the validity of our approach.  Here, we consider two such constraints, motivated by the design and/or manufacturing of the microswimmers, and we use a  modified  steepest ascent algorithm to account for these constraints.

\subsubsection{Constraints C1 and C2}

 In the first constraint C1, we assume that the  function 
$A(x)$ is bounded below and above by constant fixed values, $a$ and $b$ respectively. 

In the second constraint C2,  $A$ is bounded below (by $a$)  and the filament has a fixed volume.  Recalling that $\bar{A} = E\pi \bar{r}^4$, we  can rewrite it in dimensionless form as $A = \tilde{E} r^4$ with $\tilde{E} =\frac{E \pi}{ \omega \zeta_\perp} \l(\frac{r_0}{L}\r)^4 $, where $r_0$ is a characteristic thickness of the filament. The dimensionless volume of the filament is $V = 2\pi \int_0^1 r^2 {\rm d}x = 2\pi \int_0^1 \sqrt{A} {\rm d}x / \sqrt{\tilde{E}}$. The constraint for fixed volume, $V$, can thus be written as $\int_0^1 \sqrt{A} {\rm d}x = V\sqrt{\tilde{E}}/(2\pi):=V_0$. 

The two sets of constraints we consider are therefore
\begin{itemize}
\item (C1)  $a\leq A(x) \leq b$; or
\item (C2) $a\leq A(x)$ and $\int_0^1\sqrt{A}dx = V_0$.
\end{itemize}

\subsubsection{Numerical solution}
To solve the optimization problem with constraints C1 or C2 numerically, we need to modify the steepest ascent algorithm. For C1, a projection function is introduced to ensure that the solution at each iteration remains within the fixed bounds $a$ and $b$. To address C2, the projection function is adapted to enforce the condition $a\leq A(x)$, while the objective function is adjusted to incorporate the fixed volume constraint. Detailed descriptions of the algorithms used to handle both constraints are provided in appendix~\ref{sec:appedix}.

\begin{figure}[t]
\includegraphics[width=0.5\textwidth]{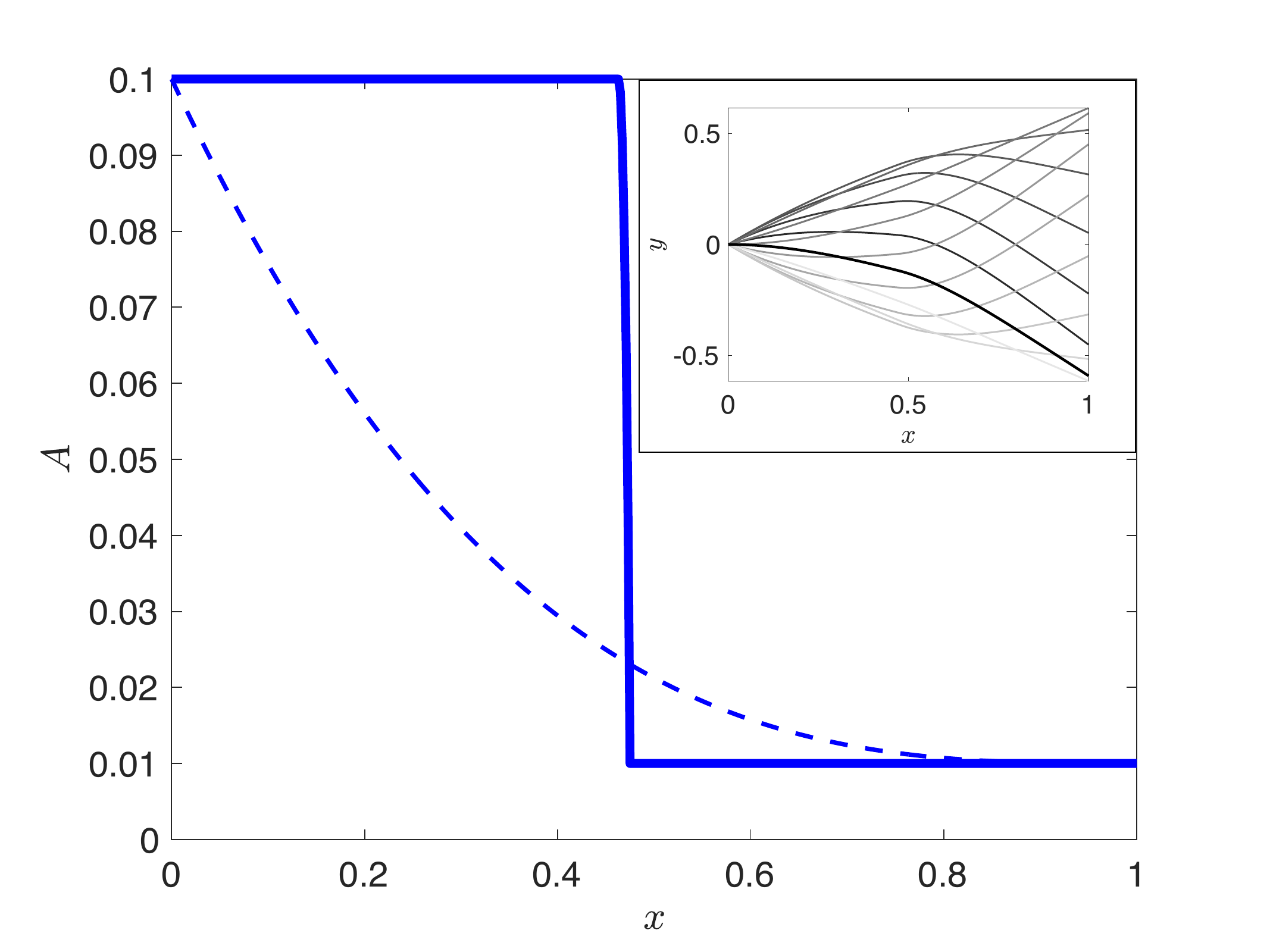}
\caption{Typical numerical solution of the optimisation problem when the dimensionless   rigidity is required to remain within the interval    $A\in [0.01,0.1]$ (bounded constraint C1). Initial condition for the function $A(x)$: dashed line; final optimal function $A(x) $: solid line. Inset: time-varying position of the centreline of the filament $y(x,t)$ at different times (greyscale) for the optimal $A(x)$.}\label{fig:num1}
\end{figure}


\section{Optimal shapes with constraint C1 (bounded bending rigidity)}\label{sec:res1}
In this section, we  use numerical simulations and theoretical analysis to determine the filament shapes that lead to optimal propulsion. Here we consider the shapes that follow the constraint C1 and thus whose  bending rigidity must remain within a prescribed  interval. 

\subsection{Optimal shapes have piece-wise constant rigidity}

We first run numerical simulations for  several chosen values of $[a,b]$ and for an initial smooth profile satisfying the boundary conditions $A_1(0)=b$, $A_1(1)=a$ with $A_1'(1)=A_1''(1)=0$. An example where the shape is initially chosen to have a cubic profile is shown in Fig.~\ref{fig:num1} as  dashed line in the case   $[a,b]=[0.01,0.1]$, an interval that contains the optimal constant value from Fig.~\ref{fig:const-A}.

We  applied the algorithm described in  \S\ref{sec:constr-i} setting the tolerances to $\delta_1 = \delta_2 = 10^{-5}$. Independently of the shape taken for the initial condition, we systematically find that the final solution for the optimisation problem is a step-like function, illustrated with  solid line in Fig.~\ref{fig:num1}, that takes value $A=b$ at the proximal end, the value of $A=a$ at the distal end and undergoes a sudden jump between $a$ and $b$ at a dimensionless point $x_0$; for example, we have $x_0\approx 0.47$ for $[a,b]=[0.01,0.1]$. 
When the filament is built of the same material throughout, a step-like function for the bending rigidity corresponds to a piecewise-constant radius for  the filament.  The corresponding spatial positions of the filament $y(x,t)$ at different time points are shown in the inset of the Fig.~\ref{fig:num1}. The resulting propulsive force is $F \approx 0.116$, which is almost twice higher than the optimal force in the case of uniform rigidity (Fig.~\ref{fig:const-A}). 

To test whether it is the only optimal solution for given $[a,b]$, we run the algorithm for 20 random initial conditions (represented as a truncated Fourier series with random modes) and always obtain the same final step-like solution.  
For different choices of $a$ and $b$, the optimal solutions are also invariably step functions, and the values of the jump points $x_0$ depend on the values of both $a$ and $b$.

\begin{figure} [t]
\includegraphics[width=0.5\textwidth]{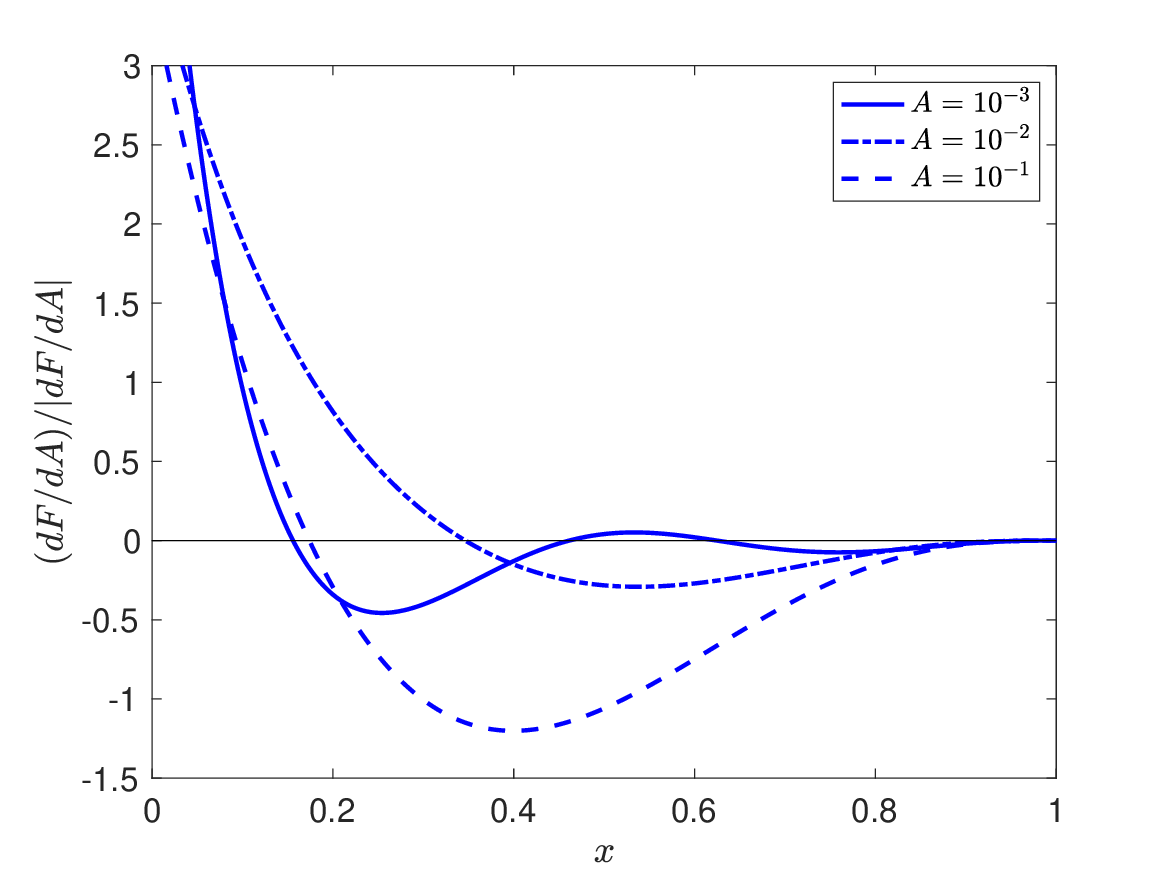}
\caption{Normalised functional derivative ${\rm d}F/{\rm d}A$ in the case of a constant dimensionless bending rigidity   $A$ (three different values, see inset). 
}\label{fig:const-A2}
\end{figure}
\subsubsection{Interpretation of the optimal solution}

An intuitive mathematical explanation for this optimal solution can be gained by examining  the shape of the functional derivative in the case of spatially uniform bending profile, $A$. When $A$ is   constant, the adjoint problem in eqs.~\eqref{eq:adj-problem}-\eqref{eq:adj-bc} can be solved analytically and we can calculate the functional derivative of $F$ with respect to $A$ exactly; we show the results in Fig.~\ref{fig:const-A2} for different dimensionless values of $A$.  Clearly, the derivative 
${\rm d}F/{\rm d}A$ is always positive at $x=0$, so in the optimisation steps we add positive values to $A$ around $x=0$ until we reach the upper bound $b$.  On the other hand, near $x=1$ the functional derivative is always negative, and this drives the repeated reduction of the bending stiffness $A$ near the distal end of the filament with each iteration step until we reach the lower bound $a$. 

Physically, our solution indicates that the front of the optimal  filament tends to be stiff, which creates a finite amplitude for the beat (akin to a filament with clamped end). The distal part, in turn, is much softer, which leads to an asymmetry in the stroke and generates propulsion.

\subsection{Analytical solution for piece-wise constant $A$}
The numerical results suggest that the optimal bending profile takes the form of a piece-wise constant value. For this type of profile, we can in fact compute a fully analytical solution for the filament shape $y(x,t)$. 

Assuming that $A$ takes form $A = b + (a-b)H(x-x_0)$, where $H$ is a Heaviside function (i.e.~$H(x) = 0$ for $x\leq 0$ and 1 for $x>0$), we  can split the problem in equation \eqref{eq:force_balance_dimless} into two separate equations on the left and right of the jump point $x_0$
\begin{subeqnarray}
\pd{y_1}{t}=-b\pd{^4y_1}{x^{4}}, \quad x\leq x_0,  \\
\pd{y_2}{t}=-a\pd{^4y_2}{x^{4}}, \quad x>x_0. \label{eq:force_balance_dimless-pw_const}
\end{subeqnarray}
The boundary conditions at the point of discontinuity of $A$ are continuity of function $y$ and its derivative, i.e.
\begin{equation}\label{eq:bc_pw_c1}
y_1(x_0,t)=y_2(x_0,t), \quad \pd{y_1}{x}(x_0,t)=\pd{y_2}{x}(x_0,t),
\end{equation}
and continuity of the elastic torque and force,
\begin{equation}\label{eq:bc_pw_c2}
b\pd{^2y_1}{x^2}(x_0,t)=a\pd{^2y_2}{x^2}(x_0,t), \quad b\pd{^3y_1}{x^3}(x_0,t)=a\pd{^3y_2}{x^3}(x_0,t).
\end{equation}
These are accompanied by the boundary conditions on the right and left boundaries as in equation \eqref{eq:BCs}.

Looking for solutions of the form $y=Re(\tilde{y}e^{it})$, these equations are simplified to 
\begin{subeqnarray}
i\tilde{y}_1=-b \tilde{y}_1^{(iv)}, \quad x\leq x_0, \\
i\tilde{y}_2=-a\tilde{y}_2^{(iv)}, \quad x>x_0. \label{eq:force_balance_dimless-no-time}
\end{subeqnarray}

The general solution to equation~\eqref{eq:force_balance_dimless-no-time} is
\begin{subeqnarray}
\tilde{y}_1=C_1e^{\alpha x} + C_2 e^{-\alpha x}+C_3 e^{i\alpha x}+C_4 e^{-i\alpha x},\\
\tilde{y}_2=D_1e^{\beta x} + D_2 e^{-\beta x}+D_3 e^{i\beta x}+D_4 e^{-i\beta x},
\end{subeqnarray}
where $\alpha=\frac{1}{\sqrt[4]{b}} \l(\cos\frac{\pi}{8}-i\sin\frac{\pi}{8}\r)$ and $\beta=\frac{1}{\sqrt[4]{a}}\l(\cos\frac{\pi}{8}-i\sin\frac{\pi}{8}\r)$. Applying the boundary conditions from equations  \eqref{eq:bc_pw_c1}-\eqref{eq:bc_pw_c2}, we obtain a linear system for the values of $C_i$ and $D_i$, which is easily inverted numerically. 

The propulsive force~\eqref{eq:F_prop_y} can then be computed analytically using the dynamics in equation \eqref{eq:force_balance_dimless-pw_const}, boundary conditions~\eqref{eq:bc_pw_c1}-\eqref{eq:bc_pw_c2} and integration by parts,

\begin{eqnarray}
    &&F = -\frac{1}{2\pi}\int_0^{2\pi}\int_0^{1}\pd{y}{t}\pd{y}{x}{\rm d}x{\rm d}t = \nonumber \\
    &&\frac{1}{2\pi}\int_0^{2\pi} \l( b\int_0^{x_0}\pd{^4y_1}{x^4}\pd{y_1}{x}{\rm d}x +a\int_{x_0}^{1}\pd{^4y_2}{x^4}\pd{y_2}{x}{\rm d}x\r){\rm d}t   = \nonumber \\
    &&\frac{1}{4} \l(\l.b\pd{^2\tilde{y_1}}{x^2}\pd{^2\tilde{y}^*_1}{x^2}\r|_{x=0} -b\l.\l(\pd{\tilde{y}_1}{x}\pd{^3\tilde{y}^*_1}{x^3}+\pd{\tilde{y}^*_1}{x}\pd{^3\tilde{y}_1}{x^3} \r)\r|_{x=0}\r.+\nonumber \\
   && \l. \l.\frac{b}{a}(b-a)\pd{^2\tilde{y}_1}{x^2}\pd{^2\tilde{y}^*_1}{x^2}\r|_{x=x_0}\r), \label{eq:f_prop_const}
\end{eqnarray}
where we used stars to denote complex conjugate.

\subsection{Maximum propulsive force for different bounds $[a,b]$}

   \begin{figure}[t]
\includegraphics[width=0.45\textwidth]{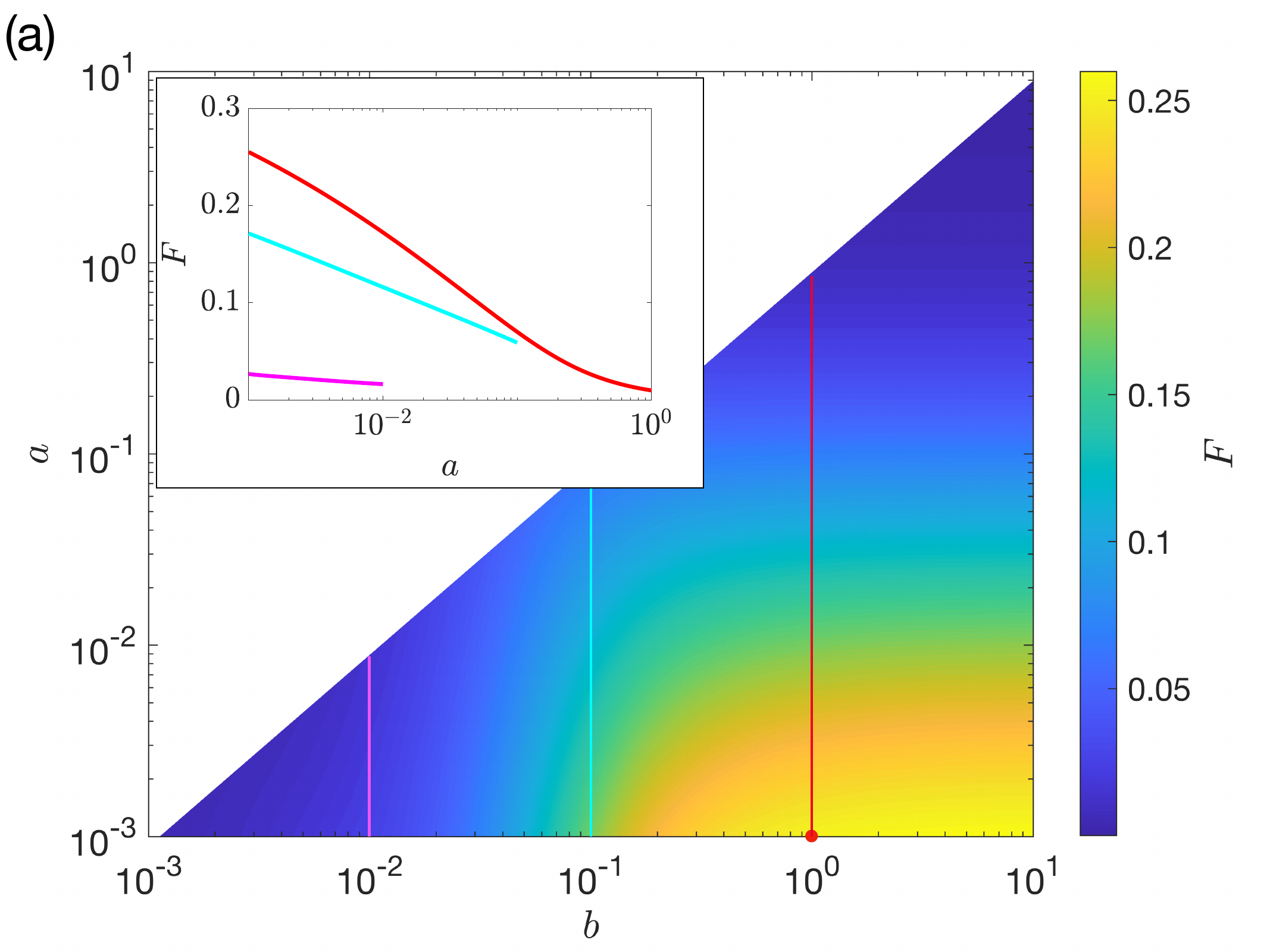}\\
\includegraphics[width=0.45\textwidth]{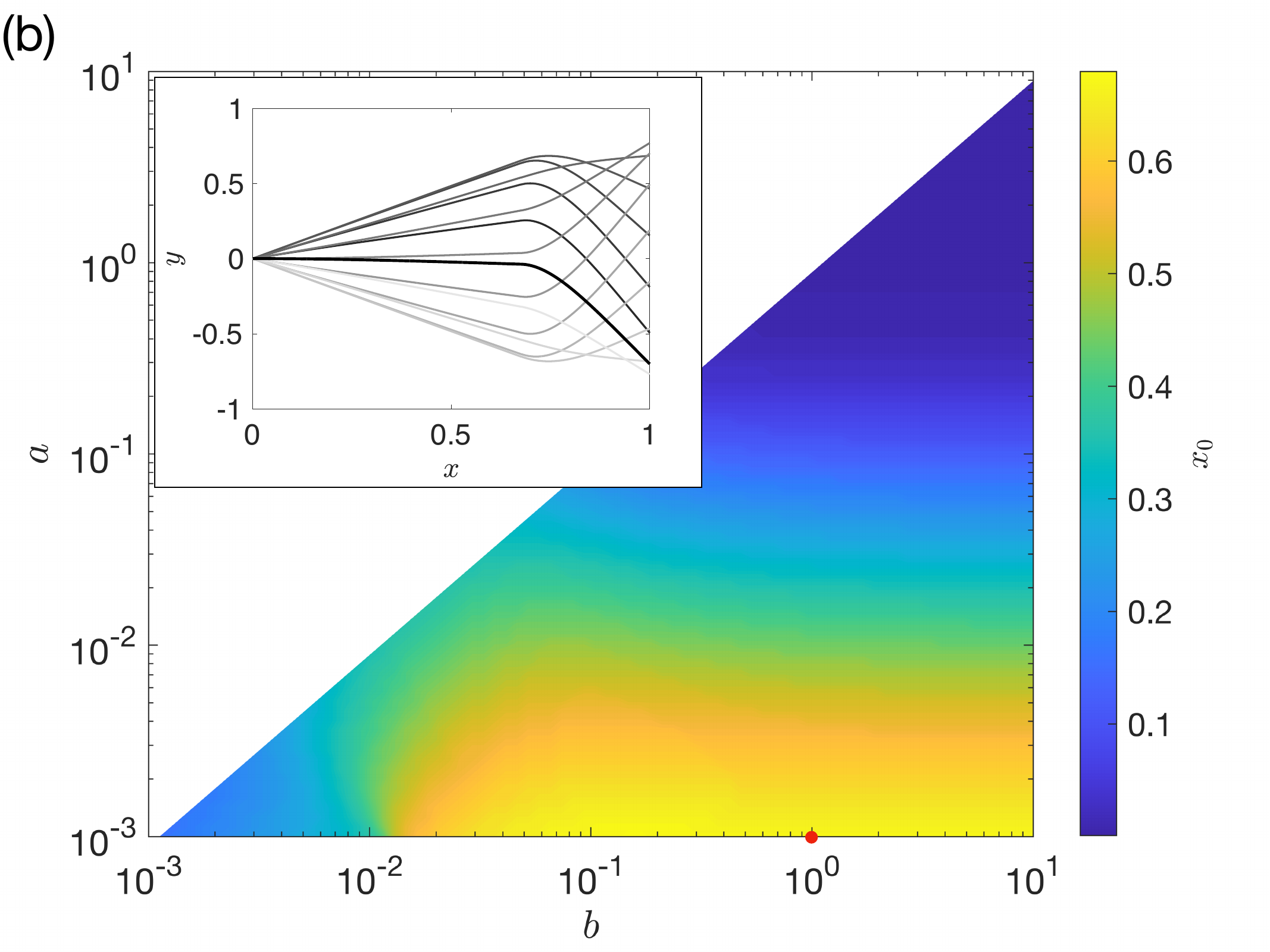}
\caption{Optimisation of a filament with piecewise constant bending rigidity $A$. (a) Maximum dimensionless $F$ (colours) as a function of dimensionless interval limits $a$ and $b$.  Inset: maximum value of $F$ for fixed $b$ and variable $a$. The lines correspond to the lines of the same colour in the main plot: red is for $b=1$, cyan for $b=0.1$ and magenta for $b=0.01$. (b) Corresponding values of the optimal transition point $x_0$. Inset: the position of the centreline of the filament $y(x)$ at different times for $a=10^{-3}$ and $b=1$ (denoted by a red dot on the plot); grey scale corresponds to different times during the period of oscillation (starting with thicker black line and fading to lighter grey colours).}\label{fig:maxF}
\end{figure}

The existence of this analytical solution makes it straightforward to mathematically study the impact of varying the values of the bounding constraints $a$ and $b$ on the propulsive force. In Fig.~\ref{fig:maxF}a we show the dependence of the maximum propulsive force $F$ on different dimensionless values of $a$ and $b$, both varying from $10^{-3}$ to 10. The corresponding value for the  jump point $x_0$ is displayed in Fig.~\ref{fig:maxF}b.

We first observe from Fig.~\ref{fig:maxF}a  that $F$ grows systematically when $b$ is increasing and  $a$ is decreasing. Increasing the value of $b$ further than 1, however, results in a very slow growth of $F$. Indeed, for $a=10^{-3}$, increasing $b$ from $b=1$ to $b=10$, results in increasing $F$ by only about 2\%. Our results show that $F$ reaches a maximum of $\approx 0.26$, which is over four times higher than the maximum value of 0.06 obtained with a spatially uniform  rigidity (see Fig.~\ref{fig:const-A}).

\subsubsection{Limiting behaviour for large $b$ and small $a$}

In the limit of large  $b$ and low  $a$ (i.e.~the bottom right corners of Fig.~\ref{fig:maxF}, and thus close to the unbounded limit), the optimal values for $x_0$ appear to all converge to $x_0\approx 0.67$, as shown in Fig.~\ref{fig:maxF}b. To understand the physical origin of this limiting value, we may consider separately the contribution of the rigid and soft parts of the filament to the propulsion.

\begin{figure}[t]
 \centering
\includegraphics[width=0.33\textwidth]{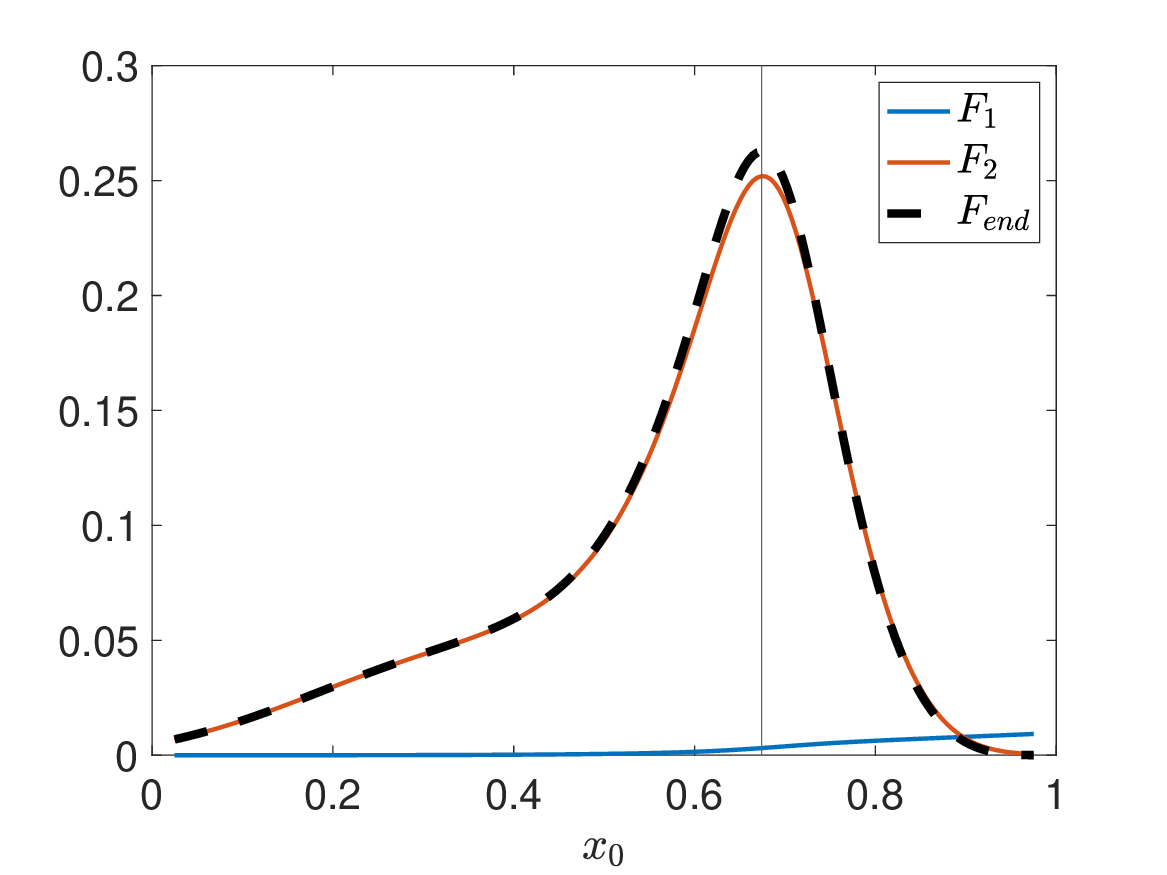}
\caption{Dimensionless propulsive force as a function of $x_0$ for $a=0.001$ and $b=1$. $F_1$ and $F_2$ correspond to the contribution of the proximal and the distal ends of the filament, respectively, with $F_1:=F-F_2$. $F_{end}$ corresponds to a simplified model, where the proximal end is considered rigid. The vertical line corresponds to the optimal location $x_0\approx 0.67$.}\label{fig:F1_F2}
\end{figure}

 The contribution of the distal end of the filament to the propulsion ($x>x_0$) is captured by the term 
\begin{equation}
F_2:=\frac{b}{4a^2}\l.\pd{^2\tilde{y}_1}{x^2}\pd{^2\tilde{y}^*_1}{x^2}\r|_{x=x_0}=\frac{a}{4}\l.\pd{^2\tilde{y}_2}{x^2}\pd{^2\tilde{y}^*_2}{x^2}\r|_{x=x_0},
\end{equation}
in   equation \eqref{eq:f_prop_const} (where we have used the condition \eqref{eq:bc_pw_c2} for the second equality), and the contribution from the proximal end is then defined as $F_1:=F-F_2$. In Fig.~\ref{fig:F1_F2} we plot the values of  $F_1$ and $F_2$ for $a=0.001$ and $b=1$ (shown as red dots in Fig.~\ref{fig:maxF}) as functions of $x_0$. For intermediate values of $x_0$ (away from $0$ and $1$), the distal part of the filament ($x>x_0$, $F_2$, red line) gives the dominant contribution to the propulsive force, while the almost rigid proximal part ($x<x_0$, $F_1$, blue line) is always much smaller. This suggests that the optimal value of $x_0$ is determined by  the dynamics of the soft (distal) part of the filament.

To understand physically why  the optimal value of $x_0$ is set by the flexible   end, let us consider a simplified setup and focus solely on the distal end portion, $x\in [x_0,1]$, in the limit where it is set into motion by an oscillating  proximal end that is straight and rigid; in that case, by reversibility, the proximal end  does not contribute to any propulsion. The problem in equation~\eqref{eq:force_balance_dimless-pw_const}   then reduces to,
\begin{eqnarray}\label{eq:37}
\pd{y_2}{t}=-a\pd{^4y_2}{x^{4}}.
\end{eqnarray}    
The position and slope of this filament at $x=x_0$ have to match with those of the rigid (proximal) piece of the filament~\eqref{eq:bc_pw_c2}. Thus, the position will follow the displacement at point $x_0$, $y(x_0,t) = x_0\sin t$ and the slope will coincide with the actuation at the base of the rigid filament $y_x(x_0,t) = \sin t$; the actuation of the flexible portion of the filament is therefore a combination of  oscillating position and slope.   

The solution to equation~\eqref{eq:37} can be found analytically, as it is a problem for a filament with constant bending rigidity.\cite{wiggins1998flexive} The resulting propulsive force, termed $F_{end}$, is shown in Fig.~\ref{fig:F1_F2} as a function of $x_0$ for the values  $a=0.001$ and $b=1$. The optimal value for  $x_0$ is seen to be the same for this simplified problem  as for the full optimisation, demonstrating that the propulsion is indeed governed by the distal portion of the filament.  In this simplified problem, the location of $x_0$ determines both the length of the soft distal part of the filament ($1-x_0$) and the displacement at its base. The intermediate value of $x_0$ is set by a balance between the requirement of having a finite-sized length for the oscillating distal part (so $x_0$ not too close to $1$) and the requirement of having a finite oscillating amplitude (so $x_0$ not too close to $0$). 




\section{Optimal shapes with constraint C2 (fixed filament volume)}\label{sec:res2}

The previous section demonstrated that, in the case of filaments whose bending rigidities are constrained to remain within a prescribed interval (constraint C1), the optimal filament shape was always piece-wise constant. We now consider the optimisation problem with the fixed-volume constraint  (C2).

\subsection{Optimal shapes have smoothly-decreasing bending rigidity, followed by a  constant value}
 
\begin{figure}[h!]\centering
\includegraphics[width=0.45\textwidth]{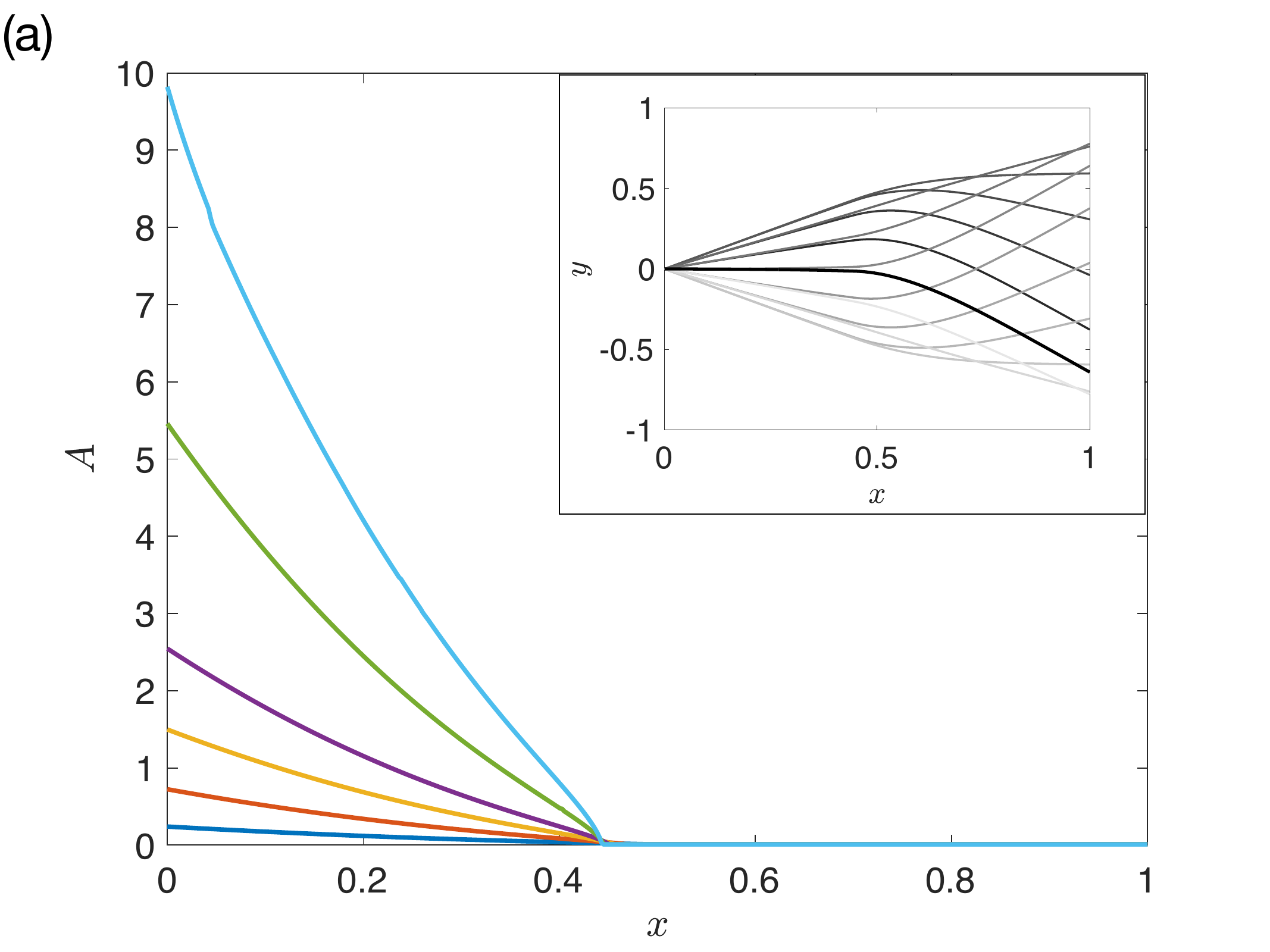}
\includegraphics[width=0.45\textwidth]{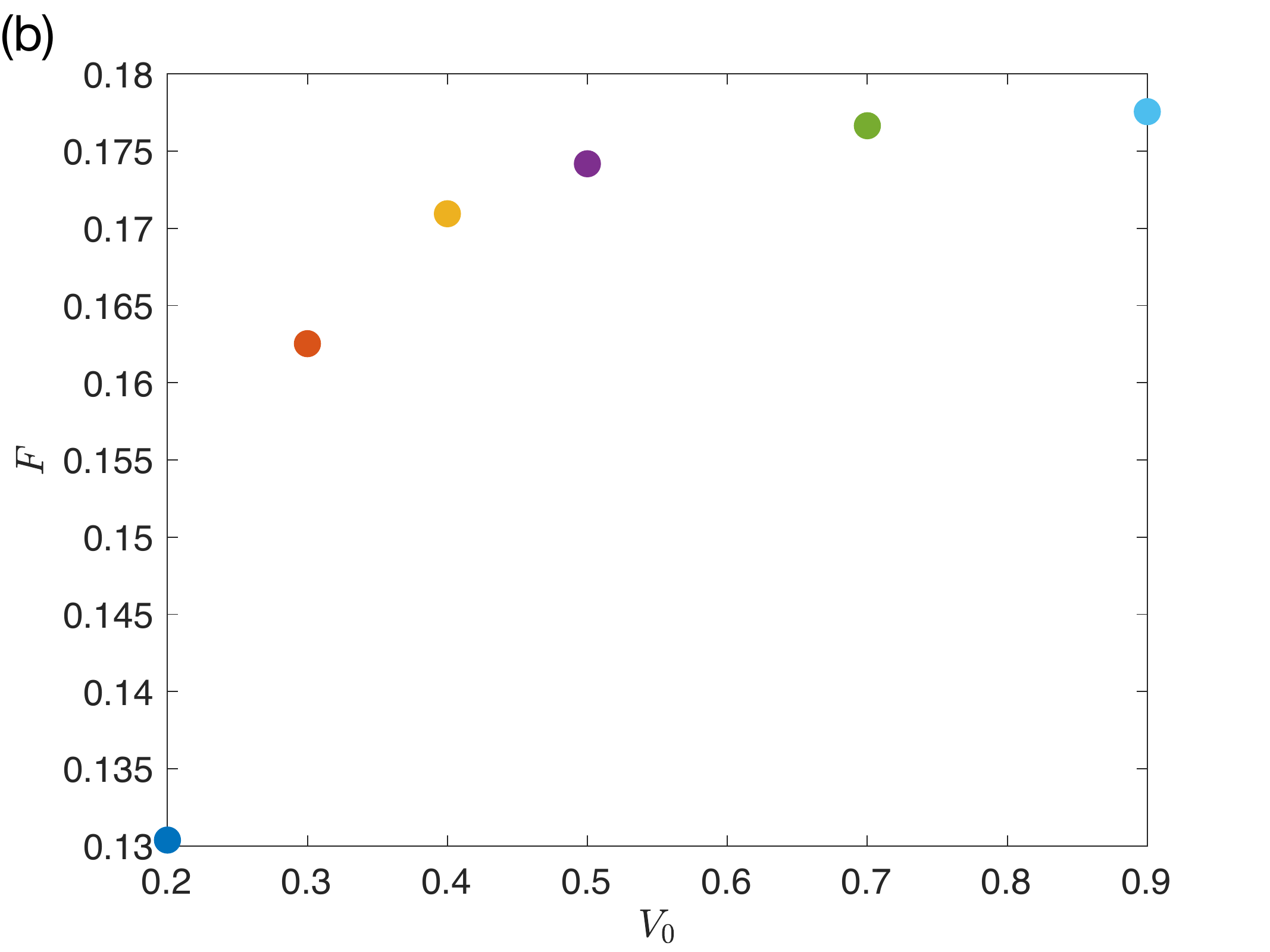}
\caption{Optimisation of a filament with of fixed volume. 
(a) Numerical solution for optimal $A$ for various values of fixed volume and $a=0.01$; the colours correspond to different values of $V_0$. Inset: Position of the filament $y(x)$ at different time points for the optimal $A$ for $V_0 = 0.5$ (purple curve in the main figure). Black thick line corresponds to $t=0$ and further times are shown with fading greyscale. 
(b)  The  dimensionless propulsive force $F$ corresponding to each optimisation
as a function of $V_0$ ($x$-axis) and with $a=0.01$; the colours are the same as in (a). 
}\label{fig:fixed_V}
\end{figure}

We  use the algorithm in appendix \ref{sec:constr-ii} to solve the optimisation problem  for given values of $a$ and $V_0$ starting with a random initial condition. The resulting optimal bending functions $A$ for different values of $V_0$ and fixed $a=0.01$ are shown in Fig.~\ref{fig:fixed_V}a; the inset shows the time-varying shape of the filament for the optimal profile in the specific case where $V_0=0.5$ (purple curve in main figure).

Similarly in the bounded case, the value of $A$ is seen to be constant and equal to the lower boundary $a$ beyond  a certain location $x_0$. However, in contrast with the previous case, closer to the base (i.e~ for $x < x_0$) the bending rigidity $A$ is no longer constant but is always  a continuously decreasing function of $x$.

In Fig.~\ref{fig:fixed_V}b we show the resulting propulsive force. We find that larger prescribed volumes $V_0$ lead to systematically higher propulsive force. However, as seen in  the figure, the increase slows down;  from $V_0=0.4$ to $V_0=0.9$, the value $F$ increases only by about 4\%, while from $V_0=0.7$ to $V_0=0.9$ only by 0.5\%. 

In the case where the material has a constant Young's modulus, our result for the optimal bending rigidity means we have a solution for  optimal radius of the filament, given by  $r=(A/\tilde{E})^{1/4}$. An example of the optimal $r$ for $V=0.5$ and $a=0.01$ is shown with solid red line in the inset of Fig.~\ref{fig:A-sqrt}. 


\subsection{Approximated optimal $A$} 
\begin{figure}[t]
\includegraphics[width=0.45\textwidth]{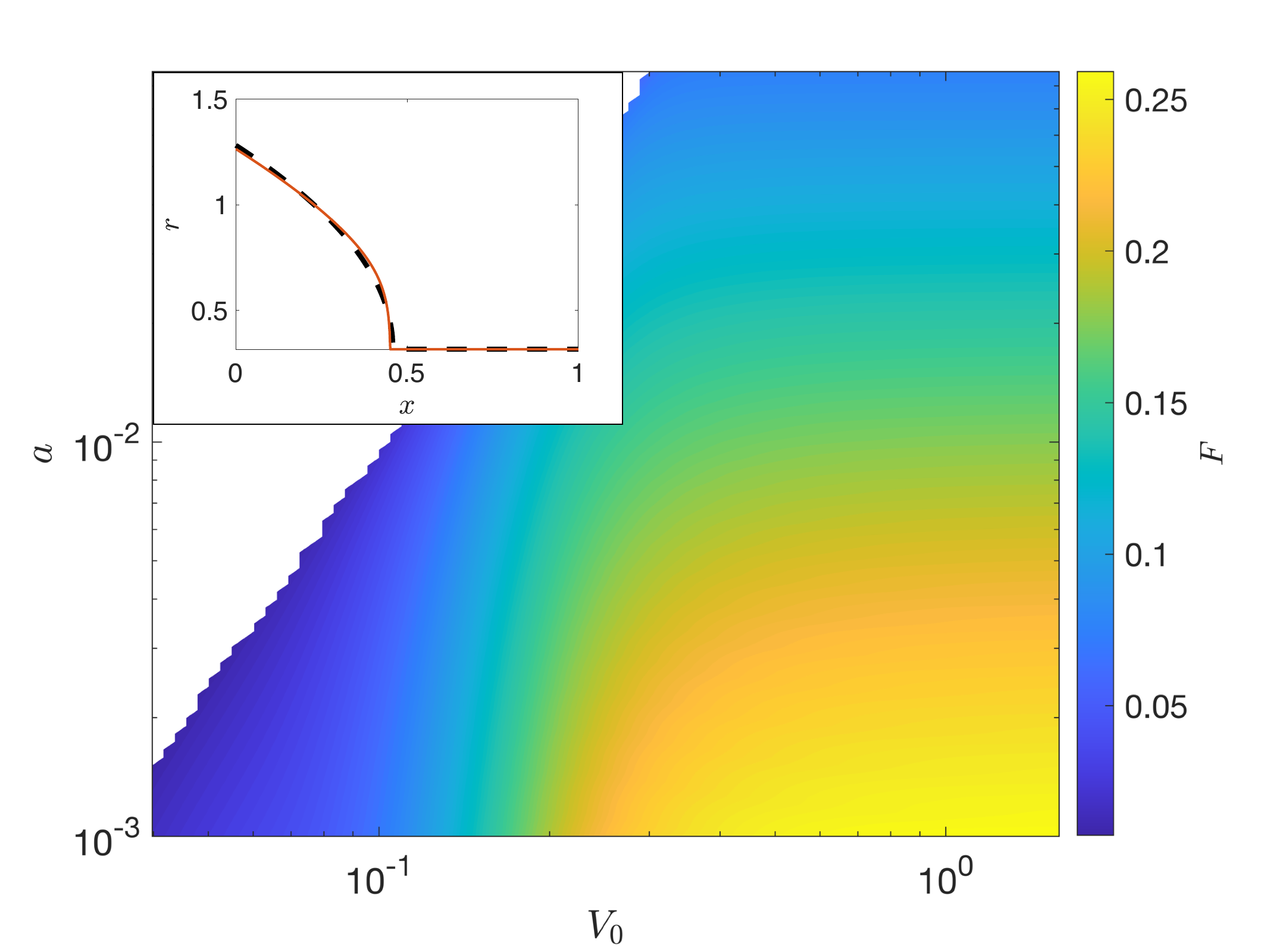}
\caption{Maximum propulsive force (colours) for $A$ given by the empirical ansatz in equation~\eqref{eq:A-sqrt} as a function of the parameters $V_0$ and $a$. Inset: The spatially variable thickness of the filament $r = \sqrt[4]{A}$ for $V_0=0.5$ and $a=0.01$ (i.e.~we assumed $\tilde{E}=1$). The solid curve is the optimal shape obtained numerically, while the dashed curve is the approximated optimal shape from  equation~\eqref{eq:A-sqrt}.}\label{fig:A-sqrt}
\end{figure}

In order to further study the dependency of $F$ on values of $a$ and $V_0$, we introduce the empirical approximation for the optimal $A$ as
\begin{eqnarray}\label{eq:A-sqrt}
A = \left\{ \begin{array}{ll}
         \l(a_0 \sqrt{x_0-x}+\sqrt[4]{a}\r)^4, & \mbox{if $x\leq x_0$},\\
        a & \mbox{if $x>x_0$},\end{array} \right. 
\end{eqnarray}
where the constant $a_0$ is chosen to satisfy the constant volume condition and where the transition point $x_0$ is fixed.  The radius of the filament, given by  $r=(A/\tilde{E})^{1/4}$ with $A$ from the model in equation~\eqref{eq:A-sqrt} is shown in the inset of Fig.~\ref{fig:A-sqrt} in black dashed line for the choices $V_0=0.5$, $a=0.01$ and $\tilde{E}=1$. The relative error for this approximation compared to the optimal case (solid red line), is below 1\% in terms of the propulsive force $F$.

Using this model, we may vary the values of both $a$ and $V_0$ and compute the maximum value for $F$ over possible values of $x_0$, as shown in Fig.~\ref{fig:A-sqrt}. The results are   very similar to those obtained with the optimisation under the constraint C1. Indeed, the propulsive force is systematically   larger if $a$ is small and if $V_0$ is large, as was seen in Fig.~\ref{fig:maxF}a, while the values of the optimal transition point $x_0$, after which the function $A$, is a constant are very similar to Fig.~\ref{fig:maxF}b (not shown).

The physical interpretation of the results in this section is very similar to the  one we proposed in the previous section and for both constraints $A$ is stiff at the base and soft at the distal end with $x_0$ governing the length of the soft distal end.

\section{Discussion}\label{sec:disc}

\subsection{Summary}

Enhancing the propulsion of elastic filaments through optimised design is a significant challenge within the realm of micro-robotics. \cite{nelson2010microrobots} One promising approach involves the optimisation of the  bending rigidity along the length of the filament, here denoted by $A$ in its dimensionless version.  In this study, we focus on the  optimisation of $A$ with the aim of maximising the propulsive force generated by a model system: an elastic slender filament with an oscillating slope at its base. To accomplish this, we employ techniques of functional optimisation, first by computing the functional derivative of $A$ using an adjoint problem, and then by using this derivative within a steepest ascent algorithm operating in the functional space. Our optimisation process considers two sets of additional constraints:  ensuring that $A$ remains within defined bounds, or  maintaining a fixed volume for the filament.

 Our main finding demonstrates that, in all tested cases, the optimal filament  must be stiff at the base and soft at the distal end for optimal propulsion, which we were able to interpret 
physically. In the case where $A$ is constrained to remain bounded between prescribed values $a$ and $b$, the optimal solution is invariably piecewise constant: $A$ assumes the upper bound near the filament's base and the lower bound at the distal end.    When $b$ is substantially larger than $a$ the propulsive force increases by more than a factor of 4 above the optimal value for a spatially uniform $A$. The transition point, $x_0$, where the rigidity abruptly goes from the upper bound to the lower bound is located about 2/3 along the filament's length, which we demonstrated  is governed by the propulsive physics at the distal end.  For the fixed volume constraint, the optimal filament continues to have  higher stiffness at the base and becomes softer  towards its distal end, but in this case the shape decreases smoothly near the base.



\subsection{Practical considerations}

The two constraints we considered indicate possible practical ways of designing an optimal elastic filament. The bending rigidity, $A$, of a filament is determined by the product of the Young's modulus of the material and the moment area of inertia, proportional to $r^4$, where $r$ is the filament's radius. Changes in $A$ may thus be obtained either by changing the Young's modulus or by modifying the filament radius. 
In the case of bounded $A$, the optimal piecewise-constant rigidity could be achieved by merging two segments with different Young's moduli. For the fixed volume case, on the other hand, the Young's modulus is fixed and spatially variable radius of the filament is given by $r=(A/\tilde{E})^{1/4}$, shown for example in the inset of Fig.~\ref{fig:A-sqrt}. 
 

To further illustrate the advantage of using a filament with varying bending rigidity, consider the (mm)-scale PDMS filament from \citeauthor{williams2014self}\cite{williams2014self} with length $L=1.5$~mm, radius $r=5$~$\mu$m and Young's modulus of 3.86~MPa. Using the viscosity $\mu=1.15$~mPa$\cdot$s\cite{williams2014self} and frequency $\omega/2\pi = 3.6$~Hz, we obtain the estimate $A \approx 0.03$. Small oscillations with angle $\epsilon = 0.4$~rad of this filament yields a propulsive force of $\bar{F}\approx0.3$~nN. If instead the distal part of the bending rigidity is reduced by one order of magnitude compared to the proximal part, the force increases by almost one order of magnitude to about $\bar{F} \approx 1.7$~nN. Similarly, for a PPy filament with $L=9$~$\mu$m, $r=$100~nm,\cite{jang2015undulatory} a Young's modulus of 100~MPa\cite{madden2007polypyrrole} and oscillating with frequency  $\omega/2\pi = 20$~Hz,~\cite{jang2015undulatory} we estimate  $A\approx 1$ in 65\% glycerol solution ($\mu=15$~mPa$\cdot$s\cite{segur1951viscosity}).  Under an  oscillation  slope of  $\epsilon = 0.4$~rad, we obtained a small propulsive force of $\bar{F}\approx0.23$~pN; however, lowering the  bending rigidity of the distal part by two orders of magnitude increases the force by a factor of about 18, to $\bar{F}\approx4$~pN.

\subsection{Comparison with past work}
Our results demonstrate rigorously the advantages of having a stiff base and a soft distal end for enhancing propulsion, compared to a uniform filament with optimal constant rigidity, which is consistent with past work. \cite{rathore2012planar,singh2017effect,kotesa2013tapered,peng2017maximizing} \citeauthor{singh2017effect}\cite{singh2017effect} reported a 22\% increase in propulsion for a linearly tapered elastic filament whose slope at the base was  oscillating, while  \citeauthor{peng2017maximizing},\cite{peng2017maximizing} found up to 12.5\% increase in propulsion using an exponential taper for a clamped oscillating filament. In  contrast, the  optimal bending rigidity reported in our work (also for imposed oscillating slope at the base)  results in  a more substantial (four-fold) increase in propulsion. 

In \citeauthor{peng2017maximizing}\cite{peng2017maximizing} the authors studied the optimal propulsion of a filament with two segments: one rigid and one soft. They showed that for the case of a clamped oscillating filament, the maximum thrust is achieved when the base is rigid and the distal end is soft, with a transition point of 0.82 of the total length. In contrast, for a hinged filament, the propulsion is maximised if the base of the filament is soft and the distal end is stiff, highlighting the importance of the actuation mechanism in determining the optimal bending rigidity. 

\subsection{Biological relevance}

We can draw parallels between our findings and the structure of a human spermatozoon. The flagellum of a spermatozoon cell is  tapered,\cite{gaffney2011mammalian} with an estimated proximal-to-distal ratio in bending rigidity of 35,\cite{neal2020doing} and after about 0.65 of the length of the filament, the distal end maintains a constant level of stiffness. Moreover, a sperm flagellum  is forced by a continuous internal actuation via molecular motors distributed along the whole flagellum~\cite{pak2015theoretical} except for the passive soft region at the distal end (the so-called `end piece'). In the study by \citeauthor{neal2020doing},\cite{neal2020doing} the authors noted that this passive end piece can result in a remarkable increase in swimming speed, up to 72\%, and a substantial boost in hydrodynamic efficiency, up to 438\%. This result is consistent with our findings, where the presence of a soft distal end is crucial for enhancing the propulsion of the filament. We note, however, that the comparison with biological cells is limited, owing to the qualitatively different actuation mechanisms.

\subsection{Outlook}

The methodology used in this study could be extended to tackle optimisation under other actuation mechanisms. This could   allow comparison with existing artificial swimmers employing  rotating~\cite{maier2016magnetic} or oscillating magnetic heads.~\cite{jang2015undulatory} When applied to   the case of a clamped filament with an oscillating base, we found much more complex optimal shapes 
than for a hinged filament, hinting at a very rich optimal landscape.

Additionally, our study focused on maximising solely the propulsive force, and it will be  valuable to explore other objectives such as maximising swimming speed or hydrodynamic efficiency. In our simplified  model, maximising swimming speed for a swimmer with a passive filament would lead to the same optimal solution. Indeed, in that case the steady horizontal constant velocity $V$ of the swimmer is obtained by balancing drag and propulsion as $V=\bar{F}/(D_h + \int_0^L \zeta_{\parallel} d\bar{x})$, where $D_h$ is drag coefficient for the head.~\cite{pak2015theoretical,singh2017effect} Since, at small amplitude,   only the propulsive force depends on filament dynamics, the optimisation for  $V$ is identical to that of $\bar{F}$.

 
Our study exhibits certain limitations that could be subject to future research. Firstly, our assumption of linearised elastohydrodynamics means that our model is not suitable for describing large displacements of the filament. Secondly, we rely on resistive-force theory, which simplifies our analysis by neglecting hydrodynamic interactions between different parts of the filament. While this is a reasonable approximation for small amplitude deformations, it may not hold true when the filament undergoes significant curvature. Furthermore, the model does not account for the change in drag coefficients 
 with $r$, which appear in the force balance equation~\eqref{eq:force_balance} and in the expression of the propulsive force~\eqref{eq:F_dim}. In the present work, we looked at changes in bending stiffness up to 3-4 orders of magnitude, which would correspond to changes in $r$ by about one order of magnitude. This will result in changing drag coefficients by about a factor of 2, which may alter the precise quantitative nature of the  optimal solution. 


\section*{Data availability}
All results and figures in the article were generated using Matlab R2024a. The Matlab codes can be found at the following DOI: \href{https://doi.org/10.5281/zenodo.15111675}{10.5281/zenodo.15111675}.

\section*{Conflicts of interest}
There are no conflicts to declare.

\section*{Acknowledgements}
The authors would also like to thank M. T\u{a}tulea-Codrean for useful discussions and  feedback. This project has received funding from the European Research Council under the European Union's Horizon 2020 Research and Innovation Programme (Grant No. 682754 to E.L.).



\appendix

\section{Appendix}\label{sec:appedix}

 \subsection{Algorithm for constraint C1 (bounded $A$)}\label{sec:constr-i} 
 For a bounded $A(x)$, we have to make sure that at each optimisation step, it remains within the chosen domain $[a,b]$. To ensure this, we introduce a projection operator $P(A)=\max(a,\min(A,b))$ and we adapt the projected gradient method, which consists of the following steps: \cite{hinze2008optimization}
\begin{itemize}
\item Choose the initial condition $A_1$, $a\leq A_1\leq b$.
\item For $k=1,2,3,...$, repeat the following steps until convergence criteria are achieved 
\begin{enumerate}
\item Solve for filament dynamics, equations \eqref{eq:no_time_problem}-\eqref{eq:no_time_problem-bc}, and the adjoint problem, equations~\eqref{eq:adj-problem}-\eqref{eq:adj-bc}, with given $A_k$. We use Matlab solver \texttt{bvp4c} to solve the ODE system;
\item Set $s_k=dF/dA \l|_{A_k} \r.$, where the functional derivative is computed using equation \eqref{eq:func-der};
\item Choose step $\sigma_k$ by projected Armijo rule (see below) such that $F(P(A_k+\sigma_k s_k))>F(A_k)$;
\item Set $A_{k+1}:=P(A_k+\sigma_k s_k)$.
\end{enumerate}
\end{itemize}
Defining the errors 
\begin{equation}
{\rm err}_1 = |F(A_k)-F(A_{k-1})|/F(A_{k-1}), \quad {\rm err}_2 = \max|A_k - A_{k-1}|,
\end{equation}
the convergence criteria are picked to be ${\rm err}_1<\delta_1$ and ${\rm err}_2<\delta_2$, where
and $\delta_1$ and $\delta_2$ are prescribed tolerances.

To choose the optimisation step $\sigma_k$, we use the projected Armijo rule and choose maximum $\sigma_k\in\{1,1/2,1/4,...\}$ for which 
\begin{equation}
F(P(A_k+\sigma_k s_k))-F(A_k)\geq \frac{\gamma}{\sigma_k}\|P(A_k+\sigma_k s_k)-A_k\|^2, 
\end{equation}
where $\|\cdot  \|$ refers to the $L_2$ norm. As shown in classical work, this algorithm converges.\cite{hinze2008optimization}

\subsection{Algorithm for constraint C2 (fixed volume)}  \label{sec:constr-ii}
In the case of constraint (C2), we use a similar algorithm to the one reported in \S~\ref{sec:algo} but modify the objective function to account for constant volume constraint, 
\begin{equation}
F_{V} = F + \rho \l| \int_0^1 \sqrt{A}\mathrm{d}x - V_0\r| .
\end{equation}
Here,   the second term $\rho$ is a penalty parameter for the volume constraint $\int_0^1 \sqrt{A}{\rm d}x - V_0=0$. The optimisation algorithm is then similar to the one   in \S~\ref{sec:constr-i},  replacing $F$ with $F_V$ and considering the projection function only for lower boundary, $P(A) = \max(a,A)$. Since we do not want the final solution to depend on the value of $\rho$, we   set the penalty parameter  $\rho$ to be an increasing sequence $\rho_m = \rho_0 2^m$, where $\rho_0$ is an initial choice. For each value of $\rho_m$ we solve the minimisation problem reported in \S~\ref{sec:constr-i} until   convergence is reached for given tolerances $\delta_1^m$ and $\delta_2^m$. For convenience, we work in terms of the radius $r$ instead of $A$, and thus the errors at each step $k$ are ${\rm err}_1 = |F(A_k)-F(A_{k-1})|/F(A_{k-1})$ and ${\rm err}_2 = \max|r_k - r_{k-1}|$. We then consider the next value of $\rho_m$ and follow the procedure again, until the error ${\rm err}_3 = |F(A_m)-F(A_{m-1})|/F(A_{m-1}) + \max|r_m - r_{m-1}|$ gets smaller than a set tolerance $\delta_3$ or a maximum number of iterations is reached. In our solution we set $\min{\delta_1^m}=10^{-4}$, $\min{\delta_2^m}= 10^{-4}$  and $\delta_3=5\cdot 10^{-5}$.


\renewcommand\refname{References}

\bibliography{Bibliography_elastic_flagellum} 
\bibliographystyle{rsc} 

\end{document}